Original Research

# Differential brain connectivity patterns while listening to breakup and rebellious songs: A functional magnetic resonance imaging study

Chia-Wei Li[1] and Chen-Gia Tsai[2,3*]

[1] *Department of Radiology, Wan Fang Hospital, Taipei Medical University, Taipei, Taiwan*

[2] *Graduate Institute of Musicology, National Taiwan University, Taipei, Taiwan*

[3] *Neurobiology and Cognitive Science Center, National Taiwan University, Taipei, Taiwan*

[*] Corresponding author: tsaichengia@ntu.edu.tw (Chen-Gia Tsai)

## Abstract

Song appreciation involves a broad range of mental processes, and different neural networks may be activated by different song types. The aim of the present study was to show differential functional connectivity of the prefrontal cortices while listening to breakup and rebellious songs. Breakup songs describe romance and longing, whereas rebellious songs convey criticism of conventional ideas or socio-cultural norms. We hypothesized that the medial and lateral prefrontal cortices may interact with different brain regions in response to these two song types. Functional magnetic resonance imaging data of fifteen participants were collected while they were listening to two complete breakup songs and two complete rebellious songs currently popular in Taiwan. The results showed that listening to the breakup songs, compared to the rebellious songs, enhanced the coupling between the medial prefrontal cortex and several emotion-related regions, including the thalamus, caudate, amygdala, hippocampus, middle orbitofrontal cortex, and right inferior frontal gyrus. This coupling may reflect the neural processes of pain empathy, reward processing, compassion, and reappraisal in response to longing and sorrow expressed by the breakup songs. Compared to the breakup songs, listening to the rebellious songs was associated with enhanced coupling between subregions in the prefrontal and orbitofrontal cortices. These areas might work in concert to support re-evaluation of conventional ideas or socio-cultural norms as suggested by the rebellious songs. This study advanced our understanding of the integration of brain functions while processing complex information.

## 1. Introduction

Complicated mental processes are engaged during attentive listening to a song. There has been growing interest in understanding the neural correlates underlying various aspects of song perception, including the effects of lyrics on musical emotion processing [1], and the integrative processing of lyrics and tunes during song encoding [2]. A few brain-imaging studies of popular songs focused on their rewarding effects, finding that the subcortical reward region was more sensitive to popular songs compared to artistic songs [3], and that the subcortical reward region worked in tandem with the auditory cortex to establish unfamiliar popular songs as desirable [4]. A majority of prior imaging studies used short song excerpts as stimuli, with systematic manipulations to engage and isolate mental processes involved in song perception.

In real life, we listen songs holistically rather than locally or separately. The complexity of semantic content of song lyrics, musical elements (e.g., melody, harmony, phrasing, mode, tempo, metric organization, syllable rhythm, groove, loudness, instrumentation, and singing style), and formal structures (e.g., verse-chorus form, bar form, or the twelve-bar blues) seems critical to listeners' enjoyment of songs. While the dynamic and complex nature of human songs could obstruct researchers' attempts to isolate and manipulate independent variables, there is an emerging trend towards more ecologically valid neuroimaging experiments [5-7]. Using naturalistic stimuli such as complete songs could enable us to study listening experiences under real-life condition, and thereby advances our understanding of the integration of specific brain functions while processing complex information.

It is worth noting that song-listening experiences differ across song types. For example, breakup songs and rebellious songs are two important types of popular songs and have different impacts on listeners. A majority of breakup songs are slow songs dealing with the loss of romantic relationships, with musical emphasis on beautiful, bittersweet melodies. Rebellious songs, in contrast, are often of an antisocial nature with emphasis on aggressive drumming. People often use breakup, rebellious, or other songs to regulate their moods, but music does not always have positive psychological impacts [8]. In particular, there is little knowledge on how breakup and rebellious songs arouse emotional and cognitive reactions. The aim of the present study was to characterise the functional connectivity patterns in the brain during listening to breakup and rebellious songs.

Analysis of brain functional connectivity has been employed to explore the neural correlates of musical emotions [9] and preference/evaluation of music [4, 10]. In the present study, we chose two seed regions in the prefrontal cortex based on previous studies. The first seed was the medial prefrontal cortex (mPFC). It has been found that the mPFC contributes to encoding subjective value of music [4, 11]. This evaluation process might be linked to mPFC's functions of emotion awareness and affective empathy [12-14]. During exposure to sad music, compared to happy music, the dorsal and ventral clusters in the mPFC were found to have significantly higher centrality values (i.e., higher importance of network nodes) [9]. Given that breakup songs often convey sorrow and longing, the mPFC is likely to play an integrative role during listening to breakup songs.

On the other hand, the rebellious songs used in the present study contain messages of self-questioning and criticizing conventional socio-cultural norms. We hypothesized that the appreciation of rebellious songs may relate to *cognitive flexibility*, which refers to the ability and willingness to adjust one's emotional, cognitive, and behavioural responding to a situation based on new information [15]. A classic task used to assess cognitive flexibility is *reversal learning*, in which subjects need to adjust behaviour when reinforcement contingencies change [16] or when someone explicitly gives the novel rules [17]. The latter type of reversal learning, instructed reversal learning, has recently attracted considerable attention in brain research. Here we draw parallels between the criticism of conventional socio-cultural norms in rebellious songs with the verbal instruction in a reversal learning task. For example, a rebellious song used in the present study mentioned 'I pursue a kind of wealth, yet must disperse with all I have.' This statement may instigate the participants to reverse the association between wealth and reward long taken for granted. Reversal learning entails several subprocesses supported by different brain regions. The striatum and orbitofrontal cortex (OFC) may be responsible for the processing of changing reinforcement contingencies [18-20], while the rostral prefrontal cortex (rPFC) seems to integrate rules into a unified task set [21]. Since the rPFC plays a top-down control role in cognitive flexibility, we selected the left rostrolateral prefrontal cortex (rlPFC) as the second seed region for functional connectivity analysis.

Two main hypotheses were examined in the current study. Given that breakup songs often deal with painful longing, we hypothesized that the mPFC increasingly interact with the regions implicated in the processing of pain, emotion, and compassion, compared to rebellious songs. Moreover, we hypothesized that the left rlPFC increasingly interact with the OFC to support re-evaluation of conventional ideas or socio-cultural norms while listening to rebellious songs compared to breakup songs. In this study, results for functional magnetic resonance imaging (fMRI) data analysis were combined with the result of a questionnaire about subjective listening experiences to provide putative explanations for differential functional connectivity patterns in response to breakup and rebellious songs.

## 2. Materials and methods

### 2.1. Participants

Participants were recruited via a public announcement on the internet. Sixteen healthy adults participated in the present study. The experimental data in this study were obtained with the informed consent of all participants. The Research Ethics Committee of the National Taiwan University approved the research protocol, code No. 201409HM016. The research protocol was in compliance with the Declaration of Helsinki. According to self-reports, no participants had a history of neurological or psychiatric disorders. One participant withdrew from the study due to a mild claustrophobia issue. Therefore, the final sample analysed included 15 participants (female/male = 8/7; 21–26 years-old).

### 2.2. Stimuli and procedure

The musical stimuli involved two breakup songs ('Betrayal', https://youtu.be/xHsSWiLsIRY; 'A lost shoal', https://youtu.be/Ie1KcGvBN_k) and two rebellious songs ('2012', https://youtu.be/YygVVvuIWm4; 'Seeing the mountain upon opening the door', https://youtu.be/F-V4zb_LAno). These verse-chorus songs were sung in Mandarin and popular in Taiwan a few years ago. All four songs were extracted from commercially available compact discs and converted into digital files (44,100-Hz sampling rate, 16-bit stereo). The intensity of all four songs was normalised so that the maximum sound level was approximately 95 dB. Stimuli were presented through scanner-compatible headphones. Additionally, participants wore earplugs to reduce the scanner noise by 20–30 dB.

A pre-scan test was conducted to make sure that stimuli were familiar to the participants. This test involved a 10-second auditory presentation of each stimuli, after which participants were asked to indicate the title of this songs and to recite the lyrics of this passage. All participants passed this test with a perfect (100%) correct score. Prior to the fMRI scan, participants were asked to rate the breakup quality and rebellious quality of these four songs on a 5-point Likert scale. During the fMRI scan, these four songs were presented once in a pseudorandom order. The participants listened to these songs with their eyes closed. Thirty-second gaps of silence were inserted between adjacent songs.

Ten to twelve months after the fMRI experiment, the participants were asked to rate their agreement to six statements on a 5-point Likert scale for each song via an online questionnaire. Among 15 participants, 14 completed this questionnaire. The result of this questionnaire was expected to reveal more fine-grained subjective experiences during song listening. These statements were as follows:

*When I listen to this song, I constantly feel the singer feeling remorse and yearning for love.*
*When I listen to this song, I feel that the singer is like a good friend of mine expressing his/her feelings.*
*The beauty of the music itself is the main reason I like this song.*

*This song questions conventional ideas.*
*When I listen to this song, I re-evaluate ideas that I have taken for granted.*
*When I listen to this song, I have a feeling of rebellion and criticizing society*

### 2.3. Acquisition and preprocessing of MRI data

For imaging data collection, the participants were scanned using a 3T Bruker MRI/MRS System (Bruker, Ettlingen, Germany) and a quadrature head coil at the Interdisciplinary MRI/MRI Lab, Department of Electrical Engineering, National Taiwan University. Functional scanning involved generating a series of 2.5 mm axial slices of the region of interest which were acquired using a gradient echo planar imaging (EPI) with the following parameters: time to repetition = 2000 ms, echo time = 35 ms, flip angle = 87°, in-plane field of view = 256 × 256 mm, and acquisition matrix = 96 × 96 × 16 to cover from the top of the striatum to the inferior temporal areas (Figure 1). This was done because (1) the two seed regions were the mPFC and rlPFC, and (2) the present study focused primarily on the subcortical emotion processing systems and the evaluative mechanisms implemented in the OFC. One EPI scanning with whole-brain coverage (TR/TE = 4500/30 ms) was then acquired from each participant for spatial individual-to-template normalization in data-preprocessing. After all the fMRI scans, one rapid imaging with refocused echoes (RARE) T1-weighted imaging (TR/TE/TI = 4873/17.41/553 ms) with spatial resolution of 1 × 1 × 2.5 mm³ was acquired.

The first five volumes (silent period in those scans) were discarded to allow the magnetization to approach a dynamic equilibrium. The preprocessing of images was performed with SPM12 (Wellcome Trust Centre for Neuroimaging; http://www.fil.ion.ucl.ac.uk/spm) and Resting-State fMRI Data Analysis Toolkit (REST) [22]. For motion correction, aligning each volume to a reference base volume was performed across the functional dataset. Each functional dataset was filtered using a low-pass Chebyshev Type II filter with a frequency range of 0 to 0.1 Hz (in MATLAB; MathWorks, Inc., Natick, MA, USA). After filtering, linear trends were removed to eliminate signal drift induced by system instability. The individual functional images were coregistered with the whole-brain-coverage EPI images and were normalised to the corresponding Montreal Neurological Institute (MNI) space by applying transferred parameters, which calculated from the whole-brain-coverage EPI images and MNI template, and linearly resampled to an isotropic resolution (2 × 2 × 2 mm³). Finally, all datasets were smoothed using a 4 mm FWHM Gaussian kernel to minimize inter-individual variances and to enhance SNR.

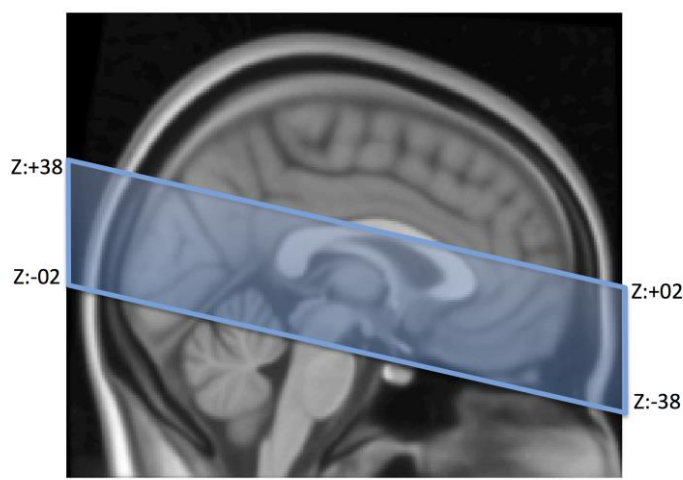

**Fig. 1. Spatial coverage of fMRI.**

*2.4. Functional connectivity analysis*

According to the a priori hypothesis based on previous imaging studies, we selected two seed regions of interest (ROIs), including the left mPFC (centred at [−2, 49, 7]) [12] and left rlPFC (centred at [−36, 48, 6]) [23]. The radius of each spherical seed ROI was set at 4 mm (each ROI contained 33 voxels). In individual data analysis, the correlation coefficient map between each seed-ROI and the whole scanned brain regions was calculated using REST toolkit, and the average of the time-series data obtained from the cerebral spinal fluid region and from the white matter were used as the covariate factors as well as the six motion-parameter covariates, which were estimated using the SPM12 software package. At the group-level analysis, the Fisher's *r*-to-*z* transformation was applied to the individual connectivity maps among the selected ROIs for the following group-level analysis, and the differences between the breakup song condition and the rebellious song condition were calculated using SPM12 software. Statistical significance was thresholded at cluster-level FDR corrected to $p < 0.05$ with a minimum cluster size of 10 contiguous voxels.

# 3. Results

Paired *t*-tests were applied to the questionnaire data for the breakup and rebellious songs. A Bonferroni-corrected threshold of $p < 0.05$ was employed. Figure 2 shows the questionnaire data for the breakup and rebellious songs. The qualities of breakup, yearning for love, and the extent of musical beauty to influence personal preference of the two breakup songs were significantly stronger/greater than the two rebellious songs. The qualities of rebellion, convention-breaking, and self-questioning of the two rebellious songs were significantly stronger than the two breakup songs. Therefore, the participants perceived the breakup songs and the rebellious songs as typical breakup songs and rebellious songs, respectively.

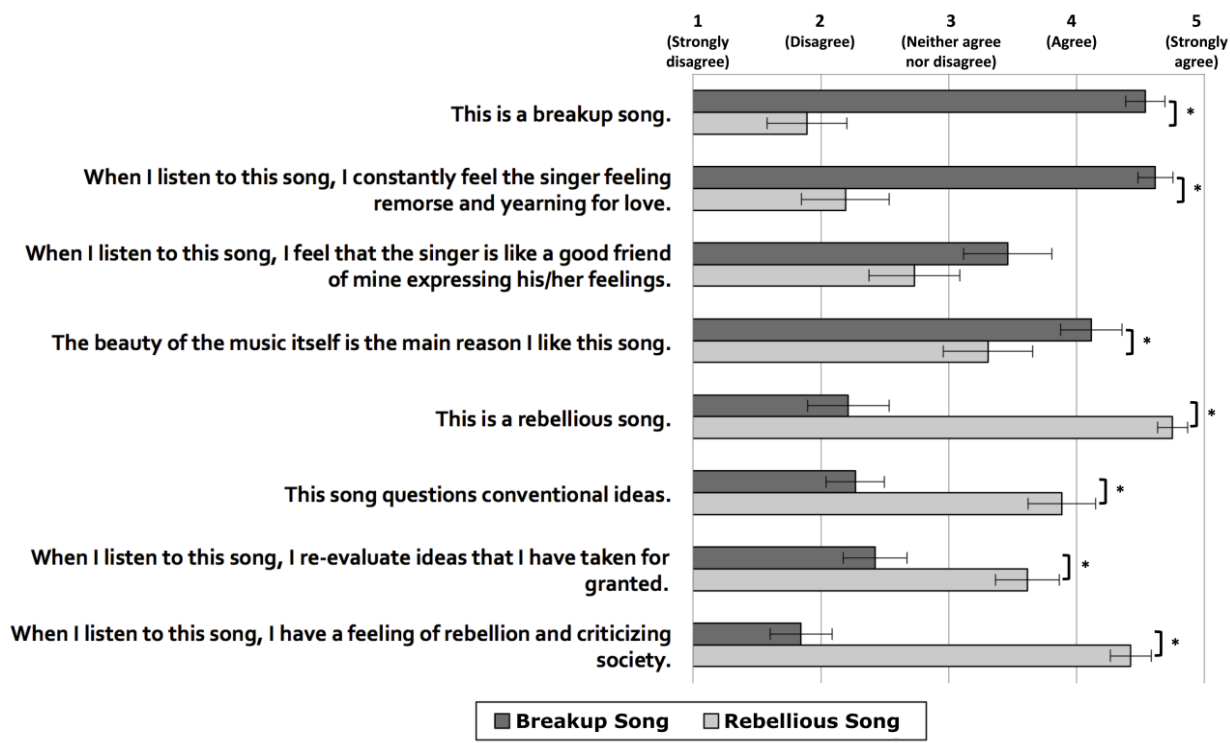

**Fig. 2. Questionnaire data showing the average ratings on a 5-point Likert scale.** Error bars indicate standard deviations.
*$p < 0.05$ (Bonferroni-corrected)

Results of functional connectivity analysis are presented in Figure 3, Figure 4, and Table 1. For the contrast of breakup minus rebellious songs, the mPFC displayed significantly increased functional connectivity with the right transverse temporal gyrus (auditory cortex), left middle OFC, medial thalamus, bilateral caudate, left amygdala, left hippocampus, and pars triangularis of the right inferior frontal gyrus (IFGtri). No significant increase of the mPFC functional connectivity for the rebellious songs compared to breakup songs was observed. For the contrast of rebellious minus breakup songs, the rlPFC displayed significantly increased functional connectivity with the right middle OFC, pars orbitalis of the right IFG (IFGorb), and right medial OFC. No significant increase of the rlPFC functional connectivity for the breakup songs compared to rebellious songs was observed.

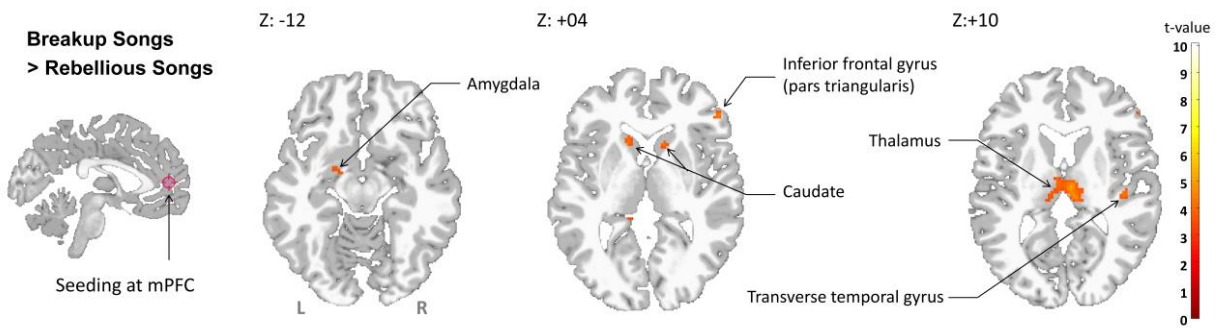

**Fig. 3. Increased functional connectivity between the mPFC and emotion-related regions for the breakup songs than for the rebellious songs** (FDR-corrected $p < 0.05$, at least 10 contiguous voxels). Abbreviate: mPFC, medial prefrontal cortex)

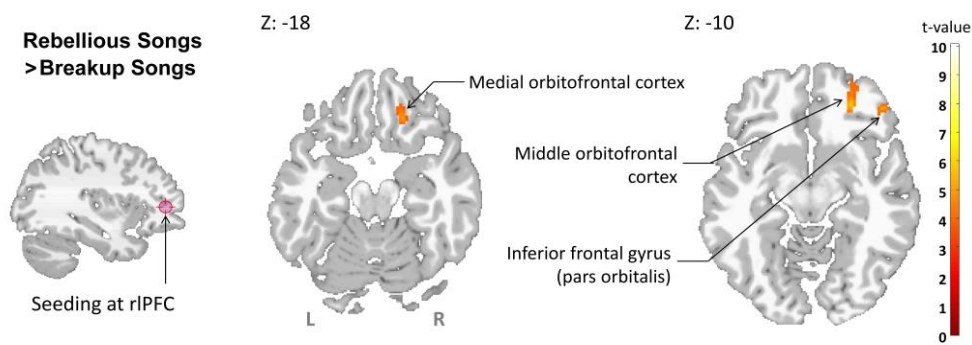

**Fig. 4. Increased functional connectivity between the rlPFC and the OFC for the rebellious songs than for the breakup songs** (FDR-corrected p < 0.05, at least 10 contiguous voxels). Abbreviate: rlPFC, rostrolateral prefrontal cortex)

**Table 1. Comparisons of the functional connectivity of two seed regions (mPFC and rlPFC) while listening to breakup and rebellious songs.**

| Volume Information | MNI Coordinate | | | t- value | Cluster | BA |
|---|---|---|---|---|---|---|
| | X | Y | Z | | (Voxel) | |
| *Seeding at mPFC (Breakup songs > Rebellious songs)* | | | | | | |
| Thalamus | 6 | -16 | 10 | 5.09 | 204 | |
| Inferior frontal gyrus (pars triangularis) | 54 | 38 | 4 | 4.58 | 23 | 45 |
| Caudate | 14 | 16 | 6 | 4.55 | 38 | |
| | -12 | 18 | 4 | 4.35 | 17 | |
| Superior/middle orbitofrontal gyrus | -30 | 54 | -4 | 4.48 | 37 | 10 |
| Transverse temporal gyrus | 44 | -22 | 10 | 4.41 | 26 | 41 |
| Hippocampus | -12 | -36 | 2 | 4.23 | 10 | |
| Amygdala | -18 | -4 | -12 | 4.06 | 13 | |
| *Seeding at mPFC (Rebellious songs > Breakup songs)* | | | | | | |
| No differential activations | | | | | | |
| | | | | | | |
| *Seeding at rlPFC (Rebellious songs > Breakup songs)* | | | | | | |
| Middle frontal gyrus, superior/middle orbitofrontal gyrus | 28 | 42 | -10 | 5.55 | 122 | 10, 11 |
| Inferior frontal gyrus (pars orbitalis), inferior orbitofrontal gyrus | 48 | 40 | -10 | 4.93 | 49 | 47 |
| Superior orbitofrontal gyrus | 20 | 34 | -18 | 4.78 | 57 | |
| *Seeding at rlPFC (Breakup songs > Rebellious songs)* | | | | | | |
| No differential activations | | | | | | |

## 4. Discussion

The current fMRI study aimed at assessing the functional connectivity patterns during song appreciation. While undergoing fMRI, participants listened to two familiar breakup songs and two familiar rebellious songs. The breakup songs use lyrical music to describe sorrow and longing, whereas the rebellious songs use music with propulsive rhythms to convey criticism of conventional socio-cultural norms. We hypothesized that these two song types may lead to different functional connectivity patterns of the mPFC and rlPFC in listeners. As expected, these two seed regions changed their functional connectivity across the two song types. The contrast of breakup minus rebellious songs demonstrated that the mPFC was mainly correlated with regions implicated in affective processing. Compared to the breakup songs, listening to the rebellious songs yielded increased functional connectivity between the rlPFC and OFC, which might underpin re-evaluation of the conventional ideas or socio-cultural norms as suggested by the rebellious songs. Since the breakup and rebellious songs differ in musical features and the semantics/content of the lyrics, it is difficult to determine precisely what role these brain regions play in song appreciation. However, we discuss plausible interpretations for their involvement in light of previous neuroscientific studies. The interpretations are speculative, but they are offered in the hope of stimulating further work on the higher brain mechanisms underlying song appreciation.

### *4.1. Functional connectivity for breakup songs*

We found that breakup songs enhanced the functional connectivity between the mPFC and auditory cortex compared to rebellious songs. It was reported that the anatomical connectivity between the mPFC and auditory cortex was correlated with individual differences in reward sensitivity to music [11]. Moreover, the increased reward value of unfamiliar popular songs was associated with increased functional connectivity of the nucleus accumbens to the auditory cortex and vmPFC [4]. In light of these studies, our finding of the alteration in the functional connectivity between the mPFC and auditory cortex might reflect the questionnaire result that the beauty of music played a more important role in the appreciation of the breakup than rebellious songs.

The mPFC is known to be engaged in social-emotional processing of information about the self [12, 13] and empathy [14]. In the current study, the mPFC showed increased connectivity with several subcortical regions for the breakup songs versus rebellious songs, including the thalamus, amygdala, and caudate. The medial thalamus has been implicated in the processing of physical pain [24, 25], pain from social rejection [26], and empathy for pain [27]. Previous studies have shown that dominance of affective empathy was associated with stronger mPFC-amygdala functional connectivity [28, 29]. The questionnaire data of the present study also show that the participants increasingly felt the singer feeling remorse and yearning for love during listening to the breakup songs compared with the rebellious songs. We suggest that the stronger mPFC-thalamus connectivity and mPFC-amygdala functional connectivity might reflect participants' increased affective empathy of painful breakup during listening the breakup songs than during listening to the rebellious songs. Alternatively, the increased mPFC-amygdala functional connectivity for the breakup songs versus the rebellious songs might also reflect positive emotion in response to the bittersweet beauty or pain relief in breakup songs, because coactivation of the mPFC and amygdala was observed during the experience of positive emotions [30]. Moreover, the increased coupling between the mPFC and left hippocampus for breakup than for rebellious songs might contribute to the processing of autobiographical memory associated with the beautiful breakup songs. Enhanced functional connectivity between the hippocampus and auditory cortex was observed during listening to a favorite song [10], and sad music tends to evoke rumination, reflectiveness, as well as nostalgia in listeners [9, 10, 31]. Unfortunately, our questionnaire did not include the question whether the participants remembered autobiographical events while listening the two breakup songs.

Previous studies have linked the caudate with reward processing and reward-related emotion experience [32, 33]. For example, coactivation of the mPFC and caudate was found during monetary gains [34]. Remarkably, up-regulation of positive affect via compassion-meditation on affective responses to depictions of individuals in distress was associated with increased caudate activity [35]. It would be interesting in future studies to determine whether listening to breakup songs facilitates the endogenous generation of positive affect in a similar manner with evidence from the regulatory mechanism of compassion-meditation.

In addition to subcortical regions, the right IFGtri and left middle OFC also displayed enhanced coupling to the mPFC for the breakup songs compared to rebellious songs. Activity in the right IFGtri was found to be greater during cognitive reappraisal of negative emotional stimuli than during watching negative emotional stimuli [36]. Moreover, this region might underpin compassion for other's suffering. The right IFGtri was found to exhibit increased activity while viewing pictures of human suffering previously verified to elicit compassion [37]. Increased activity in the right IFGtri was also observed in individuals who were successfully taking a third-person perspective while viewing pictures of hands and feet in painful situations [38]. In the present study, the enhanced mPFC-IFGtri functional connectivity while listening to the breakup songs might reflect that the participants adopted a self-distanced perspective toward breakup songs. The left middle OFC was also found to play a role in cognitive reappraisal of emotional information. Prior research indicated that the left middle OFC was involved in reappraisal of negatively-valenced stimuli [39], voluntary emotion suppression [40], and elaboration of emotional autobiographical memory [41]. We suggest that the enhanced functional connectivity between the mPFC and left middle OFC while listening to the breakup songs might reflect that the participants reappraised and/or elaborated the negative messages in the breakup songs.

Our finding of increased interactions between mPFC and the brain regions representing one's own feelings/emotions and compassion is consistent with McKlveen [42], who proposed that the mPFC may coordinate stress responses to generate context-appropriate behaviour. The present study provided preliminary evidence that aesthetic experience of breakup songs involves both negative and positive emotions. The negative emotions were due to pain empathy and likely mediated by the interactions between the mPFC, thalamus, and amygdala. On the other hand, the positive emotions might be due to (1) a detached aesthetic attitude toward the beauty of music, and (2) compassion with feelings of warmth, concern and care for the other. This finding echoes the previous finding that aesthetic appreciation and compassionate characteristics may be critical to the enjoyment of sad music [31]. Moreover, a recent model of the enjoyment of negative emotions in art reception indicated that the aesthetic virtues of artistic representation and the compositional interplays of positive and negative emotions are critical to the integration of negative emotions into altogether pleasurable trajectories [43]. Our data extend this view by demonstrating that the neural correlates of feeling positive emotion from sad songs might be the mPFC, auditory cortex, amygdala, caudate, IFGtri, and middle OFC.

### 4.2. Functional connectivity for rebellious songs

The rebellious songs used in the present study contain messages of self-questioning and criticizing conventional socio-cultural norms. The results from the questionnaire reveal that while listening to the rebellious songs, compared to the breakup songs, the participants had a stronger feeling of rebellion and criticizing society, increasingly re-evaluating ideas long taken for granted. We hypothesized that re-evaluating stereotyped viewpoints shared common neural mechanisms with cognitive flexibility, which refers to the ability and willingness to adjust one's responding to a situation based on new information [15]. Therefore, we chose a seed region in the left rlPFC, which plays a top-down role in reversal learning [21]. A number of studies on reversal learning used conditions of probabilistic feedback, in which reinforcement contingencies change unexpectedly. There has been recent interest in instructed reversal learning, in which a verbal instruction specifies the novel rule governing the stimulus-reward contingencies [17]. The criticism of conventional socio-cultural norms in rebellious songs may be analogous to the verbal instruction in an instructed reversal learning task. As shown by the questionnaire responses, the participants agreed that the rebellious songs questioned conventional ideas. Therefore, it is a reasonable conjecture that the participants tended to reverse certain stimulus-reinforcement associations while listening to the rebellious songs.

A growing body of evidence suggests that the left rlPFC plays a role in reversal learning. Ruge and Wolfensteller observed that instructed reversal learning was supported by similar regions as feedback-driven reversal learning, including the OFC and caudate [44]. Their results also showed activation in the left rlPFC (with peak at $[-39, 47, 10]$) for the contrast of reversal learning versus initial learning. Moreover, Cole and colleagues suggested that the rPFC may play a top-down control role in integrating rules into a unified task set, showing activation in the left rlPFC (with peak at $[-29, 49, 12]$) for the contrast of novel versus practiced trials [21]. These two reported peaks are close to the seed region of the left rlPFC in our study (centred at $[-36, 48, 6]$). We chose this seed region on the basis of a study by Hampshire and colleagues [23], who found that the left rlPFC and several regions in the OFC exhibited greater activation at the point of reversal of learning, compared to repetitions in the contingency change phase after receiving negative feedback.

Compared to the breakup songs, listening to the rebellious songs enhanced the coupling of the left rlPFC with three clusters in the OFC. The first was the right middle OFC, which has been implicated in reversal learning [45, 46] and goal-directed implicit learning [47]. The role of the right middle OFC in flexible learning may be detecting feedback changes and transforming them to behavioural adjustment [48], or evaluation of the outcome via confirming/updating the associations between stimuli/behaviour and rewards [49]. This view accords well with the finding that the right middle OFC mediates subjective value representations for abstract rewards but not innate rewards [50]. Compared to innate rewards, evaluation of abstract rewards may increasingly engage processes related to context-dependent decision-making and cognitive flexibility. Accumulating evidence has indicated that the right middle OFC enhances behavioural flexibility through updating valuations [51, 52]. Moreover, the right middle OFC was also found to mediate cognitive reappraisal [53]. Based on these studies, we interpret the increased coupling between the left rlPFC and right middle OFC for the rebellious songs versus breakup songs as a marker of re-evaluating conventional ideas or socio-cultural norms via updating the associations between stimuli/behaviour and rewards.

The second cluster that increasingly interacted with the left rlPFC during listening to the rebellious songs was the right IFGorb, which has been implicated in reversal learning [19], spatial working memory [54], social working memory [55], integrating the current word meaning with the preceding information [56], and integrating world knowledge during comprehension of nonliteral metonymic sentences [57]. We suggest that the increased functional connectivity between the rlPFC and right IFGorb during listening to the rebellious songs might reflect the role of the right IFGorb in unification of discourse information with relevant knowledge during comprehension of the rebellious songs.

The third cluster that increasingly interacted with the left rlPFC during listening to the rebellious songs was the right medial OFC. It was demonstrated that when adolescents were making risky decisions, activity in the right medial OFC was positively correlated to the degree to which these adolescents were influenced by the opinions of peers [58]. The increased coupling between the rlPFC and right medial OFC for the rebellious songs versus breakup songs might reflect that the participants' valuations were influenced by the convention-breaking ideas conveyed by the rebellious songs. Moreover, neuroimaging studies have implicated the medial OFC in maladaptive rumination or self-protection strategies to cope with self-criticism [59-61], and unsuccessful reappraisal to regulate anxiety [53]. In light of these studies, we speculate that the medial OFC might contribute to affective processes in response to the convention-breaking ideas conveyed by the rebellious songs.

The results of the present study only partially support our hypothesis that re-evaluating stereotyped viewpoints during listening to the rebellious songs shared common neural mechanisms with reversal learning. While coactivation of the rlPFC and OFC has been observed during reversal learning tasks, the OFC regions observed in the contrast of rebellious minus breakup songs were not completely identical to those observed in previous studies on reversal learning [21, 23, 44]. Given the complexity of rebellious songs, further research is needed to explore the relationship between listening to rebellious songs and cognitive flexibility.

### 4.3. Limitations

Some limitations of this exploratory study must be taken into account when interpreting these results. First, our fMRI data did not cover the full brain. Second, only two songs were used for each song type. Third, the relatively small sample size of participants afforded limited power for statistical analyses. Future research should collect whole-brain fMRI scans and increase the sample size of popular songs as well as participants to enhance the statistical power. The fourth limitation was that we did not dissociate the effects of lyrics and musical elements, and therefore more specific and precise interpretations of the functional connectivity patterns associated with rebellious songs require future research. This limitation was due to the fact that manipulations of song lyrics and musical elements could substantially alter participants' familiarity with the stimuli, which is critical to emotional and cognitive processing of music [62]. Future studies could use unfamiliar songs as stimuli to dissociate the effects of lyrics and musical elements.

## 5. Conclusions

The present study reported preliminary evidence of alterations in the functional connectivity pattern of two prefrontal regions during listening to different song types. The mPFC exhibited enhanced couplings with regions implicated in pain empathy, reward processing, compassion, and reappraisal for the breakup than rebellious songs. This result lends support to the notion that both negative and positive emotions are induced while listening to bittersweet breakup songs. When comparing the rebellious songs with the breakup songs, stronger functional connectivity between the rlPFC and a few subregions of the OFC was found. These regions might contribute to the cognitive and emotional adjustments during re-evaluating the conventional ideas or socio-cultural norms. Given the varied functions and forms across song types, we suggest that this pilot study needs to be expanded with other song types for a broader understanding of the higher brain mechanisms underlying song appreciation.

## Abbreviations

BA, Brodmann area; IFG, inferior frontal gyrus; IFGorb, pars orbitalis of inferior frontal gyrus; IFGtri, pars triangularis of inferior frontal gyrus; mPFC, medial prefrontal cortex; OFC, orbitofrontal cortex; rPFC, rostral prefrontal cortex; rlPFC, rostrolateral prefrontal cortex.

## Ethics approval and consent to participate

The experimental data in this study were obtained with the informed consent of all participants. The Research Ethics Committee of the National Taiwan University approved the research protocol, code No. 201409HM016.